\documentclass[aps,prl,onecolumn,superscriptaddress,groupedaddress]{revtex4}
\usepackage{amssymb}
\usepackage{color, graphicx} 
\usepackage{amsmath}
\usepackage{amsbsy} 
\usepackage{amsthm} 
\usepackage{bbm}
\usepackage{bm} 
\usepackage{float}
\usepackage{braket}
\usepackage{placeins}
\usepackage[colorlinks=true,citecolor=blue,linkcolor=red,urlcolor=red]{hyperref}
\usepackage[english]{babel}

\graphicspath{ {20 June /} }

\begin{document}

\title{A Quantum Theory for Propagation of Electromagnetic Waves through Lossy Dielectrics in Presence of Third Order Dispersion}
\author{Abolfazl Safaei Bezgabadi} \email{abolfazl\_safaei@live.com}
\affiliation{Photonics Department, Graduate University of Advanced Technology, Haft Bagh e Alavi Hwy, Mahan, Kerman, Iran}
%
%
\author{Mohammad Agha Bolorizadeh} \email{mabolori@uk.ac.ir}
\affiliation{Physics Department, Yazd University, Yazd, Iran}
\date{12 Oct. 2018}

\begin{abstract}
In this paper, we present a quantum theory for field propagation through a three dimensional dielectric when the third order dispersion and the attenuation coefficients are included. A unique Lagrangian is defined leading to the correct equation of motion and the classical Hamiltonian. It is assumed that the dielectric has a combination of inhomogeneity, dispersion and nonlinearity. By employing constraint quantization approach the final Hamiltonian is expanded in terms of properly defined annihilation and creation operators. We obtain the quantum fields (quantum photon-polaritons fields) for propagation through the dielectric in the presence of the third order dispersion and the attenuation coefficients by using these annihilation and creation operators. The number operator in the final Hamiltonian indicates the number of photon-polaritons in the medium. The nonlinear part of the Hamiltonian could be derived by defining displacement field operator in terms of annihilation and creation operators. As a simple example, the present quantum theory is applied to field propagation through a one dimensional slab.
\end{abstract}
\maketitle


\section{Introduction}
Light-matter interaction, especially its nonlinear interactions, has been at the center of attention of scientists since 1960s \cite{1} which has resulted to the developments in science and technology. Initially, the scientists tried to prevent the negative effects of nonlinear optical deficiencies \cite{2, 3}. Later, the nonlinear optical effects were used in favor of the advancement of science and technology \cite{4, 5}. Dependence of refractive index of the governing matter on the frequency of light \cite{6}, the intensity of light \cite{7, 8}, invention of metamaterials \cite{9}, multiphoton absorption \cite{10}, nonlinear scattering \cite{11}, nonlinear effect of pulsed light \cite{12}, solitons \cite{13} and optical bistabilities \cite{14} are important aspects of the developments in optics for the past 50 years of scientific and technological advances.\\
Quantum treatment of optics and nonlinear optics opened new windows in which novel phenomena were discoverd with no classical counterpart in optics and specifically in nonlinear optics \cite{15, 16, 17}. Therefore, the quantum treatment of pulse propagation in dielectrics, specifically in dispersive and/or nonlinear media, is essential for a detailed understanding of optical phenomena, if these phenomena are to be exploited to their full potential. \\
An approach, which could always be applied, is to model a dielectric in microscopic perspective. However, this method has not been applied widely due to the complicated internal structure of a dielectric. Alternatively, macroscopic theories for nonlinear dielectrics are devised and applied with minimal complexity. Bloembergen's pioneering work \cite{18} to describe the classical theory of nonlinear optics initiated expanding activity on the field.\\ 
To quantize the electromagnetic fields, the canonical variables are to be defined by applying the classical field equations in nonlinear optics. Jauch and Watson took the first step to introduce the canonical theory for the light intraction with dielectrics \cite{19}. They applied a canonical theory to study the homogeneous and linear non-dispersive medium. Nonetheless, their model was incomplete owing to the invention of laser generating the nonlinear optical phenomena in a dielectric. Several methods were devised to develop the canonical theory for a dielectric \cite{20,21,22} but none of them could quantize inhomogeneous nonlinear dispersive dielectrics. Hillery and Mlodinow \cite{23} invented a method making use of the electric displacement fields as canonical variables for the quantization of the nondispersive, nonlinear and homogeneous dielectric medium.\\
Lossless dielectric was assumed for the early quantum treatments of the nonlinear field propagation through a homogeneous dielectric \cite{24}. Drummond and his coworker considered dispersion up to second order in their quantum theory for a lossless nonlinear homogeneous dielectric \cite{25}. Their theory could introduce quantum solitons \cite{26}, which are observed in fiber-based optical system in years afterwards \cite{27}. In general, a combination of dispersion, nonlinearity and inhomogeneity exists in real dielectrics (e.g. optical fiber). Therefore, it is necessary to integrate Hillery and Mlodinow's microscopic nonlinear theory \cite{23} with the techniques describing dispersion and inhomogeneity of dielectric media. Effective Hamiltonians \cite{28}, being widely used in quantum optics, were used in most applied techniques. Therefore, to tackle complicated problems in linear and nonlinear optics, one should obtain the corresponding Hamiltonian leading to canonical quantization. The method introduced by Kennedy and Wright \cite{29} for quantization problem led to a correct axial field equation but the obtained Hamiltonian was different from the system's classical energy. For the first time, the field propagation in a dispersive nonlinear inhomogeneous dielectric including the second order dispersion was quantized by Drummond \cite{30}. \\
Carter \cite{31} extended the previous quantum theories for pulse propagation along an optical fiber making use of different Hamiltonians. The generation of quantum states have been investigated taking into account the nonlinear properties of pulse propagation through optical fiber while it was applied to quantum non-demolition experiments \cite{32}, generation of Schrodinger's cat states \cite{33}, generation of entangled photon pair \cite{34,35,36} and the parametric down conversion \cite{37}.\\
Third order dispersion coefficient and attenuation of media are neglected in the available quantum treatments of the pulse propagation through a dielectric in the literature. Due to the important role of third order dispersion coefficient \cite{4, 5} on optical phenomena, especially when second order dispersion coefficient is zero or infinitesimal, it is essential to include the third order dispersion coefficient. Also, to achieve more reasonable and accurate results, the attenuation of medium must be considered in a theory for quantization of the pulse propagation through a dielectric. Similar to some other phenomena in physics \cite{38, 39, 40}, field propagation in dielectrics, such as optical fibers, experiences noise of different sources \cite{41, 42, 43, 44}. For example supercontinuum generation \cite{45, 46} suffers up to $50 \%$ fluctuation in its temporal intensity profile, affecting its usage in many applications \cite{47, 48, 49,50,51}. As part of the noise has quantum origin, further investigation on the quantum aspects of field propagation in dielectric media is essential. \\
In this paper, implementing the Drummond's method \cite{30}, we present a quantum theory for the three dimensional field propagation in a real dielectric. Our system has a combination of inhomogeneity, dispersion and nonlinearity including third order dispersion term as well as attenuation term. Here, it is assumed that the attenuation factor is approximately constant over our considered frequency region. This theory has the ability to add the higher order dispersions and higher nonlinear terms into the governing equations. One can find the one dimensional example of this theory in the appendix.
\section{Classical Energy and Equation of Motion} \label{sec:2}
As methodological standpoint to establish a theory for quantization of the pulse propagation through a dielectric, it is necessary to define a canonical Lagrangian. Its resulted Hamiltonian and equations of motion should respectively be equal to the classical energy and the equations of motion (Maxwell's equations) of the considerd system, in order to this canonical Lagrangian be correct. Here, the classical energy and the equations of motion are given for the pulse propagation through a dielectric in presence of third order dispersion and attenuation terms.\\
We introduce the usual definitions used in the classical electrodynamics that have been also implemented by Drummond \cite{15} and Hillery \cite{16}. One can write the energy density for an electromagnetic radiation as:
\begin{equation} \label{eq1}
H=\int_{-\infty}^{t}\left[{\bf E}(t^{\prime}, {\bf x})\cdot\frac{\partial{\bf D}(t^{\prime}, {\bf x})}{\partial t^{\prime}}+{\bf H}(t^{\prime}, {\bf x})\cdot\frac{\partial{\bf B}(t^{\prime}, {\bf x})}{\partial t^{\prime}}\right]dt^{\prime} = H_{e}+H_{m}
\end{equation}
where the energy derived from the linear part of the fields is usually the largest part of the energy. The Fourier transform of each field (e.g. electric field),
\begin{equation} \label{eq2}
{\bf E}(t, {\bf x})=\int_{-\infty}^{\infty}e^{i\omega t}{\bf E}(\omega, {\bf x})d\omega, 
\end{equation}  
could be introduced in equation \ref{eq1}.
As the fields are real; i.e. ${\textbf{E}}\left(t,\textbf{x}\right)={\textbf{E}}^{\star}\left(t,\textbf{x}\right)$, ${\textbf{E}}\left(-\omega,\textbf{x}\right)={\textbf{E}}^{\star}\left(\omega,\textbf{x}\right)$ and ${\boldsymbol{\epsilon}}\left(-\omega, {\bf x}\right)={\boldsymbol{\epsilon}}^{\star}\left(\omega, {\bf x}\right)$, one can straightforwardly arrive at $\Re({\boldsymbol{\epsilon}}\left(-\omega, {\bf x}\right))=\Re({\boldsymbol{\epsilon}}\left(\omega, {\bf x}\right))$ and $\Im({\boldsymbol{\epsilon}}\left(-\omega, {\bf x}\right)=-\Im({\boldsymbol{\epsilon}}^{\star}\left(\omega, {\bf x}\right))$ where $\Re$ and $\Im$ denote the real and the imaginary patrs of the dielectric's permitivity tensor, respectively. The imaginary part of the dielectric's permittivity is the cause of attenuation. Nonetheless, we rewrite ${\bf E}(t, {\bf x})\cdot\partial{\bf D}(t, {\bf x})/\partial t$ as a product of the two Fourier integrals as: 
\begin{align}  \label{eq3}
{\bf E}(t, {\bf x})\cdot \frac{\partial{\bf D}(t, {\bf x})}{\partial t}&=i \int_{-\infty}^{\infty} d\omega^{\prime}\int_{-\infty}^{\infty} e^{i(\omega-\omega^{\prime}) t}\omega {\bf E}^{\star}(\omega^{\prime}, {\bf x})\cdot{\boldsymbol{\epsilon}}\left(\omega, {\bf x}\right)\cdot{\bf E}(\omega, {\bf x})d\omega.
\end{align}
It is assumed that the imaginary part of permittivity is frequency independent over the considered frequency region ($\Im({\boldsymbol{\epsilon}}\left(\omega, {\bf x}\right)=\Im({\boldsymbol{\epsilon}}\left({\bf x}\right))$) and also, $(\Re({\boldsymbol{\epsilon}}\left(\omega, {\bf x}\right)))^2\gg(\Im({\boldsymbol{\epsilon}}\left({\bf x}\right)))^2$. Applying a similar argument towards the magnetic part of the energy and considering the fact that magnetic permeability is independent of frequency, the relation:
\begin{align}  \label{eq4}
{\bf H}(t, {\bf x})\cdot \frac{\partial {\bf B}(t, {\bf x})}{\partial t}&=\frac{i}{\mu}\int_{-\infty}^{\infty} d\omega^{\prime}\int_{-\infty}^{\infty} e^{i(\omega-\omega^{\prime})t}\omega {\bf B}^{\star}(\omega^{\prime}, {\bf x})\cdot{\bf B}(\omega, {\bf x})d\omega
\end{align}
is derivrd.\\
To follow, one can split each part of the frequency integral of equation \ref{eq3} into equal parts. Substituting $-\omega^{\prime}$ and $-\omega$ for $\omega$ and $\omega^{\prime}$, respectively, in one of the integrals, inserting in Eq. \ref{eq1} and integrating over time to find the electric energy density, one can obtain:
\begin{align}  \label{eq5}
H_{e}&=\int_{-\infty}^{\infty} d\omega^{\prime}\int_{-\infty}^{\infty} e^{i(\omega-\omega^{\prime}) t}{\bf E}^{\star}(\omega^{\prime}, {\bf x})\cdot \frac{\left[\omega\Re\left({\boldsymbol{\epsilon}}\left(\omega, {\bf x}\right)\right)-\omega^{\prime}\Re\left({\boldsymbol{\epsilon}}\left(\omega^{\prime}, {\bf x}\right)\right)\right] }{2(\omega-\omega^{\prime})}\cdot{\bf E}(\omega, {\bf x})d\omega \nonumber\\
&+\frac{i}{2}\int_{-\infty}^{\infty} d\omega^{\prime}\int_{-\infty}^{\infty} e^{i(\omega-\omega^{\prime}) t}{\bf E}^{\star}(\omega^{\prime}, {\bf x})\cdot\Im({\boldsymbol{\epsilon}}({\bf x}))\cdot{\bf E}(\omega, {\bf x})d\omega
\end{align}  
assuming all the local fields and stored energies are initially zero at $t=-\infty$.\\
For a narrowband field at a frequency, $\omega_{0}$, there are significant contributions to the integral in Eq. \ref{eq5}, at $\omega=\omega^{\prime}+\Delta\omega\approx\pm\omega_{0}$. Note that near $\omega_{0}$, the relation:
\begin{equation}  \label{eq6} \frac{\left[\omega\Re\left({\boldsymbol{\epsilon}}\left(\omega, {\bf x}\right)\right)-\omega^{\prime}\Re\left({\boldsymbol{\epsilon}}\left(\omega^{\prime}, {\bf x}\right)\right)\right] }{(\omega-\omega^{\prime})}\approx \frac{\partial}{\partial \omega}(\omega\Re\left({\boldsymbol{\epsilon}}\left(\omega, {\bf x}\right)\right))
\end{equation}
holds. If $\partial\left(\omega\Re\left({\boldsymbol{\epsilon}}\left(\omega, {\bf x}\right)\right)\right)/\partial\omega$ varies slowly over the field bandwidth, one can obtain the time averaged linear dispersive energy for the pulse propagation through a dielectric in presence of the attenuation term for a monochromatic field at frequency $\omega$ as (for details see suplementary):
\begin{equation}  \label{eq7}
\langle\! H_{L} \!\rangle\!=\!\!\int\! \left[\frac{1}{2}{\bf E}^{\ast}(t,{\bf x})\cdot \left(\frac{\partial}{\partial \omega}(\omega\Re\left({\boldsymbol{\epsilon}}\left(\omega, {\bf x}\right)\right))\right)_{\backsim constant}\cdot{\bf E}(t,{\bf x})\!+\!\frac{i}{2}{\bf E}^{\star}(t, {\bf x})\cdot \Im\left({\boldsymbol{\epsilon}}({\bf x})\right)\cdot{\bf E}(t, {\bf x})\!+\!\frac{1}{2 \mu}\vert{\bf B}(t,{\bf x}) \vert^{2} \right]d^{3}{\bf x},
\end{equation}
which it can be written in terms of displacement fields as:
\begin{equation}  \label{eq8}
\langle\! H_{L} \!\rangle\!=\!\!\int\! \left[\frac{1}{2}{\bf D}^{\ast}(t,{\bf x})\cdot \left({\boldsymbol {\eta}}(\omega,{\bf x})\!-\!\omega \frac{\partial}{\partial\omega}{\boldsymbol {\eta}}(\omega,{\bf x}) \right)_{\backsim constant}\cdot{\bf D}(t,{\bf x})\!+\!\frac{i}{2}{\bf D}^{\ast}(t,{\bf x})\cdot {\boldsymbol {\zeta}}(\omega, {\bf x}) \cdot{\bf D}(t,{\bf x})\!+\!\frac{1}{2 \mu}\vert{\bf B}(t,{\bf x}) \vert^{2} \right]d^{3}{\bf x}
\end{equation}  
where ${\boldsymbol {\eta}}(\omega,{\bf x})=\Re\left(\left(\epsilon(\omega,{\bf x})\right)^{-1}\right)$ and ${\boldsymbol {\zeta}}(\omega,{\bf x})=\Im\left(\left(\epsilon(\omega,{\bf x})\right)^{-1}\right)$. According to the assumptions, $\Im({\boldsymbol{\epsilon}}\left(\omega, {\bf x}\right)=\Im({\boldsymbol{\epsilon}}\left({\bf x}\right))$ and $(\Re({\boldsymbol{\epsilon}}\left(\omega, {\bf x}\right)))^2\gg(\Im({\boldsymbol{\epsilon}}\left({\bf x}\right)))^2$, ${\boldsymbol {\zeta}}(\omega,{\bf x})$ is approximately constant over the frequency region.\\
Making use of a dual potential function \cite{15, 16}, ${\boldsymbol \Lambda}$, let's define the electric displacement
and magnetic fields, respectively, as:
\begin{equation}   \label{eq9}
{\bf D}(t,{\bf x})={\boldsymbol \nabla}\times {\boldsymbol \Lambda}(t, {\bf x}) 
\end{equation}
and
\begin{equation}  \label{eq10}
{\bf B}(t,{\bf x})=\mu \dot{{\boldsymbol \Lambda}}(t, {\bf x})
\end{equation}
for 
\begin{equation}   \label{eq11}
{\boldsymbol \Lambda}(t, {\bf x})=\sum_{\nu =-N}^{+N}{\boldsymbol \Lambda}^{\nu}(t, {\bf x}) 
\end{equation}
where $2N+1$ narrow-band field number (mode number), $\nu$, is included in the field. Expansion of ${\boldsymbol {\eta}}\left(\omega,\bf{x}\right)$ up to the third order in Taylor series near the narrow-field frequency, $\omega^{\nu}$, leads to: 
\begin{equation}  \label{eq12}
{\boldsymbol {\eta}}(\omega,{\bf x})\approx{\boldsymbol {\eta}}_{\nu}({\bf x})+\omega {\boldsymbol {\eta}}_{\nu}^{\prime}({\bf x})+\frac{1}{2}\omega^{2} {\boldsymbol {\eta}}_{\nu}^{\prime\prime}({\bf x})+\frac{1}{6}\omega^{3}{\boldsymbol {\eta}}_{\nu}^{\prime\prime\prime}({\bf x})+O\left((\omega-\omega^{\nu})^{4}\right)
\end{equation}
where ${\boldsymbol {\eta}}_{\nu}({\bf x})$, $ {\boldsymbol {\eta}}_{\nu}^{\prime}({\bf x})$, ${\boldsymbol {\eta}}_{\nu}^{\prime\prime}({\bf x})$ and ${\boldsymbol {\eta}}_{\nu}^{\prime\prime\prime}({\bf x})$ are defined in suplementary.
The higher order derivatives of expansion \ref{eq12} are neglected with respect to the second order (if non-zero) and the third order terms in many applications of optical fibers. Hence, the expansion \ref{eq12} is valid for nearly all applications. It can easily be shown that the term ${\boldsymbol {\eta}}_{\nu}^{\prime}({\bf x})$ does not appear in the linear dispersive energy, Eq. \ref{eq8}.\\
Using the slowly varying envelope approximation \cite{30} and the superposition principle over narrow-band field numbers $-N$ to $N$, the average linear dispersive energy for a wideband field could be expressed in terms of the local fields and their time derivatives at the frequency $\omega^{\nu}$ as: 
\begin{align}  \label{eq13}
\langle H_{L} \rangle &\!=\!\frac{1}{12}\!\sum_{\nu=-N}^{+N}\!\int\! \left[6({\boldsymbol\nabla}\!\times\!{\boldsymbol \Lambda}^{-\nu})\cdot\left({\boldsymbol {\eta}}_{\nu}({\bf x})+i{\boldsymbol {\zeta}}_{\nu}({\bf x})\right)\cdot({\boldsymbol\nabla}\!\times\!{\boldsymbol \Lambda}^{\nu}) -3 ({\boldsymbol\nabla}\!\times\!\dot{{\boldsymbol \Lambda}}^{-\nu})\cdot {\boldsymbol {\eta}}_{\nu}^{\prime\prime}({\bf x})\cdot({\boldsymbol\nabla}\!\times \!\dot{{\boldsymbol \Lambda}}^{\nu})\right. \nonumber\\ 
& \left. -i({\boldsymbol\nabla}\!\times\!{\dddot{\boldsymbol \Lambda}}^{-\nu})\cdot {\boldsymbol {\eta}}_{\nu}^{\prime\prime\prime}({\bf x})\cdot({\boldsymbol\nabla}\!\times\!{\boldsymbol \Lambda}^{\nu}) +i({\boldsymbol\nabla}\!\times\!{\boldsymbol \Lambda}^{-\nu})\cdot {\boldsymbol {\eta}}_{\nu}^{\prime\prime\prime}({\bf x})\cdot({\boldsymbol\nabla}\!\times\!{\dddot{\boldsymbol \Lambda}}^{\nu})+6 \mu {\dot{\boldsymbol \Lambda}}^{-\nu}\cdot\dot{\boldsymbol \Lambda}^{\nu}\right]d^{3}{\bf x},
\end{align} 
where the time and the position vector dependence of the dual potential function are omitted to shorten the last equation. The nonlinear part of energy should also be considered as nonlinear features of dilecterics have acquired strong momentum for a vast number of research. Physically, the medium response functions are frequency dependent. However, in the case that response time is relatively fast, one can neglect the frequency dependence of the medium nonlinear response functions.\\
Here, the Maxwell's equations are the equations of motion for the pulse propagation through a dielectric. ${\boldsymbol\nabla}\times{\bf D}=0$ and ${\boldsymbol\nabla}\times{\bf H}=\dot{{\bf D}}$ are satisfied by definition \ref{eq9} and \ref{eq10}, respectively. Also, it is understood that the dual potential must be a transverse field as ${\boldsymbol\nabla}\cdot {\bf B}=0$. So the main equation of motion is ${\boldsymbol\nabla}\times{\bf E}=\dot{{\bf B}}$ where ${\bf E}$ could be generally given by \cite{1, 5,  15}:
\begin{align}   \label{eq14}
{\bf E}(t,{\bf x})&=\sum_{n>0} \left[\int_{0}^{\infty}{\boldsymbol {\eta}}^{(n)}(\tau_{1},\cdots,\tau_{n}, {\bf x}) \vdots
 \left({\bf D}(t-\tau_{1},{\bf x}){\bf D}(t-\tau_{2},{\bf x}) \cdots {\bf D}(t-\tau_{n},{\bf x})\right) d\tau_{1}\cdots d\tau_{n}\right]
\end{align}
where ${\boldsymbol {\eta}}^{(n)}$ is the n$^{th}$-order nonlinear response of the medium. It should be noted that ${\boldsymbol {\eta}}$ in Eq. \ref{eq12} is the real part of the linear response of the medium and, therefore, it is equal to ${\boldsymbol {\eta}}^{(1)}$. However, the quantities ${\boldsymbol {\eta}}^{(n)}$ having $n>1$ are assumed to be real and independent of frequency. Nonetheless, the total energy can be written as:
\begin{subequations}
\begin{equation}
H_{T}=H_{L}+H_{NL}.
\end{equation} 
By implementing the Hillery's method \cite{16} and using the nonlinear polarization term in Maxwell's equations \cite{1, 5}, the nonlinear part of energy
is obtained as: 
\begin{equation}           \label{eq15}
H_{NL}(t,{\bf x})= \sum_{n>1}\frac{1}{n+1}\int {\bf D}(t,{\bf x})\cdot {\boldsymbol {\eta}}^{(n)}({\bf x})\vdots\overbrace{{\bf D}(t,{\bf x}){\bf D}(t,{\bf x})\cdots{\bf D}(t,{\bf x})}^\text{n}d^{3}{\bf x}.
\end{equation}
\end{subequations}
As our goal is to define a proper canonical Lagrangian, here, its resulted equation of motion should be equal to the equation of motion for the present system which is written in terms of dual potential for a mode number, $\nu$, as: 
\begin{align}  \label{eq16}
-\mu \ddot{\boldsymbol \Lambda}^{\nu}(t,{\bf x}) &={\boldsymbol\nabla} \times \left(({\boldsymbol {\eta}}_{\nu}({\bf x})+i{\boldsymbol {\zeta}}_{\nu}({\bf x})) \cdot \left[{\boldsymbol\nabla}\times{\boldsymbol \Lambda}^{\nu}(t,{\bf x})\right]+i{\boldsymbol {\eta}}_{\nu}^{\prime}({\bf x}) \cdot \left[{\boldsymbol\nabla}\times \dot{\boldsymbol \Lambda}^{\nu}(t,{\bf x})\right]-\frac{1}{2}{\boldsymbol {\eta}}_{\nu}^{\prime\prime}({\bf x}) \cdot \left[{\boldsymbol\nabla}\times \ddot {\boldsymbol \Lambda}^{\nu}(t,{\bf x})\right] \right. \nonumber\\ 
&\left.-\frac{i}{6}{\boldsymbol {\eta}}_{\nu}^{\prime\prime\prime}({\bf x}) \cdot \left[{\boldsymbol\nabla}\times \dddot {\boldsymbol \Lambda}^{\nu}(t,{\bf x})\right]+\sum_{n>1}\sum_{\nu_{1},\cdots,\nu_{n}}{\boldsymbol {\eta}}^{({n})}({\bf x})\vdots\left(({\boldsymbol\nabla}\times{\boldsymbol \Lambda}^{\nu_1}(t,{\bf x}))\cdots({\boldsymbol\nabla}\times{\boldsymbol \Lambda}^{\nu_n}(t,{\bf x}))\right)\right)
\end{align}
by applying the slowly varying envelope approximation. Here, the slowly varying envelope approximation requires that $e^{-i\omega^{\nu}\tau}{\boldsymbol{\Lambda}}^{\nu}(t-\tau, {\bf x})$ is treated as a slowly varying envelope function of $\tau$ and it can be expanded in a Taylor series near $\tau=0$. Generally, a term proportional to ${\boldsymbol {\eta}}_{\nu}^{\prime}\left(\bf{x}\right)$ does not appear in the linear dispersive energy expression of \ref{eq13}. However, a term proportional to ${\boldsymbol {\eta}}_{\nu}^{\prime}\left(\bf{x}\right)$ appears in the wave equation as a result of changes in phase velocity due to dispersion. It should be noted that when the terms proportional to ${\boldsymbol {\eta}}_{\nu}^{\prime\prime\prime}\left(\bf{x}\right)$ and ${\boldsymbol {\zeta}}_{\nu}\left(\bf{x}\right)$ are neglected, the equations \ref{eq13} and \ref{eq16} are similar to the corresponded equations in reference \cite{30}.
\section{Canonical Lagrangian and Hamiltonian functions}      \label{sec:3}
In order to establish a quantum theory for the pulse propagation through a nonlinear dispersive dielectric in the presence of the third order dispersion and the attenuation, we define a canonical Lagrangian leading to equation of motion \ref{eq16} and the Hamiltonian equivalent to $H_{T}$. Here, the dual potential, ${\boldsymbol \Lambda}^{\nu}$, is a kind of gauge and it could be considered a transverse field which is similar to the choice of Coulomb gauge used to quantize the electromagnetic field in free space. However, it is important to note that, since ${\boldsymbol \Lambda}^{\nu}$ is different from the vector potential, the choice of gauge is not exactly identical to the Coulomb gauge. As it is obvious, the Lagrangian must be a function of the dual potential and its time derivatives. To obtain the equation of motion from a Lagrangian, these equations are reduced to their transverse form using the transverse Euler-Lagrange equations,
\begin{equation}   \label{eq17}
\frac{d}{dt}\frac{\partial L}{\partial \dot {\Lambda}_{k}^{\perp}}-\frac{\partial L}{\partial{\Lambda}_{k}^{\perp}}=0. 
\end{equation}
For the present system, we derive a proper form for the linear and nonlinear parts of the Lagrangian as (for details see suplementary):
\begin{align}  \label{eq18}
L_{L} &=\frac{1}{2}\sum_{\nu=-N}^{+N}\int \left[-({\boldsymbol\nabla}\times{\boldsymbol \Lambda}^{-\nu})\cdot ({\boldsymbol {\eta}}_{\nu}({\bf x})+i{\boldsymbol {\zeta}}_{\nu}({\bf x}))\cdot({\boldsymbol\nabla}\times{\boldsymbol \Lambda}^{\nu})-\frac{i}{2}\left(({\boldsymbol\nabla}\times {\boldsymbol \Lambda}^{-\nu})\cdot {\boldsymbol {\eta}}_{\nu}^{\prime}({\bf x})\cdot({\boldsymbol\nabla}\times \dot{{\boldsymbol \Lambda}}^{\nu}) \right. \right.  \nonumber\\ 
&\left. \left. -({\boldsymbol\nabla}\times {\boldsymbol \Lambda}^{\nu})\cdot {\boldsymbol {\eta}}_{\nu}^{\prime}({\bf x})\cdot({\boldsymbol\nabla}\times \dot{{\boldsymbol \Lambda}}^{-\nu}) \right)-\frac{1}{2}({\boldsymbol\nabla}\times \dot{{\boldsymbol \Lambda}}^{-\nu})\cdot {\boldsymbol {\eta}}_{\nu}^{\prime\prime}({\bf x})\cdot({\boldsymbol\nabla}\times \dot{{\boldsymbol \Lambda}}^{\nu}) \right.\nonumber\\ 
&\left. +\frac{i}{3}\left(({\boldsymbol\nabla}\times \ddot{{\boldsymbol \Lambda}}^{-\nu})\cdot {\boldsymbol {\eta}}_{\nu}^{\prime\prime\prime}({\bf x})\cdot({\boldsymbol\nabla}\times \dot{{\boldsymbol \Lambda}}^{\nu}) 
-({\boldsymbol\nabla}\times \ddot{{\boldsymbol \Lambda}}^{\nu})\cdot {\boldsymbol {\eta}}_{\nu}^{\prime\prime\prime}({\bf x})\cdot({\boldsymbol\nabla}\times \dot{{\boldsymbol \Lambda}}^{-\nu})\right)  \right. \nonumber\\ 
&\left. -\frac{i}{6}\left(({\boldsymbol\nabla}\times {{\boldsymbol \Lambda}}^{-\nu})\cdot {\boldsymbol {\eta}}_{\nu}^{\prime\prime\prime}({\bf x})\cdot({\boldsymbol\nabla}\times \dddot {{\boldsymbol \Lambda}}^{\nu})-({\boldsymbol\nabla}\times {{\boldsymbol \Lambda}}^{\nu})\cdot{\boldsymbol {\eta}}_{\nu}^{\prime\prime\prime}({\bf x})\cdot({\boldsymbol\nabla}\times \dddot{{\boldsymbol \Lambda}}^{-\nu})\right)  +\mu {\dot{\boldsymbol \Lambda}}^{-\nu}\cdot\dot{\boldsymbol \Lambda}^{\nu}\right]d^{3}{\bf x}
\end{align}
and
\begin{equation}           \label{eq19}
L_{NL}= -\sum_{n>1}\sum_{\nu, \nu_{1},\cdots,\nu_{n}}\frac{1}{n+1}\int ({\boldsymbol\nabla}\times{\boldsymbol \Lambda}^{\nu}({\bf x},t))\cdot{\boldsymbol {\eta}}^{(n)}({\bf x})\vdots\left(({\boldsymbol\nabla}\times{\boldsymbol \Lambda}^{\nu_1}({\bf x},t))\cdots({\boldsymbol\nabla}\times{\boldsymbol \Lambda}^{\nu_n}({\bf x},t))\right)d^{3}{\bf x},
\end{equation}
where the total Lgrangian can be written as $L_{T}=L_{L}+L_{NL}$. The linear part of Lagrangian, Eq. \ref{eq18}, includes the third order dispersion and attenuation. The canonical momenta, the equation of motion and the Hamiltonian will, respectively, be:
\begin{align}  \label{eq20}
{\boldsymbol \Pi}^{\nu}\!=\!\sum_{k}\frac{\partial L}{\partial{\dot{\Lambda}}_{k}^{\perp}}&\!=\!\frac{1}{2}\left[{\boldsymbol \nabla}\!\times\! \left(-\frac{i}{2}{\boldsymbol {\eta}}_{\nu}^{\prime}(\!{\bf x}\!)\cdot({\boldsymbol\nabla}\!\times\!{{\boldsymbol \Lambda}}^{-\nu})\!-\!\frac{1}{2}{\boldsymbol {\eta}}_{\nu}^{\prime\prime}(\!{\bf x}\!)\cdot ({\boldsymbol\nabla}\!\times\!\dot{{\boldsymbol \Lambda}}^{-\nu}) \!+\!\frac{i}{3} {\boldsymbol {\eta}}_{\nu}^{\prime\prime\prime}(\!{\bf x}\!)\cdot({\boldsymbol\nabla}\!\times\! \ddot{{\boldsymbol \Lambda}}^{-\nu})\right)\!+\!\mu \dot{{\boldsymbol \Lambda}}^{-\nu}\right]\!,
\end{align}
\begin{align}  \label{eq21}
-\mu \ddot{\boldsymbol \Lambda}^{\nu}({\bf x},t) &= {\boldsymbol\nabla} \times \left(({\boldsymbol {\eta}}_{\nu}({\bf x})+i{\boldsymbol {\zeta}}_{\nu}({\bf x})) \cdot \left[{\boldsymbol\nabla}\times{\boldsymbol \Lambda}^{\nu}({\bf x},t)\right]+i{\boldsymbol {\eta}}_{\nu}^{\prime}({\bf x}) \cdot \left[{\boldsymbol\nabla}\times \dot{\boldsymbol \Lambda}^{\nu}({\bf x},t)\right]-\frac{1}{2}{\boldsymbol {\eta}}_{\nu}^{\prime\prime}({\bf x}) \cdot \left[{\boldsymbol\nabla}\times \ddot {\boldsymbol \Lambda}^{\nu}({\bf x},t)\right] \right. \nonumber\\ 
&\left.-\frac{i}{6}{\boldsymbol {\eta}}_{\nu}^{\prime\prime\prime}({\bf x}) \cdot \left[{\boldsymbol\nabla}\times \dddot {\boldsymbol \Lambda}^{\nu}({\bf x},t)\right]+\sum_{n>1}\sum_{\nu_{1},\cdots,\nu_{n}}{\boldsymbol {\eta}}^{(n)}({\bf x})\vdots\left(({\boldsymbol\nabla}\times{\boldsymbol \Lambda}^{\nu_1}({\bf x},t))\cdots({\boldsymbol\nabla}\times{\boldsymbol \Lambda}^{\nu_n}({\bf x},t))\right)\right)
\end{align}
and
\begin{align}   \label{eq22}
H_{T}&=\frac{1}{2}\sum_{\nu=-N}^{+N}\int \left[({\boldsymbol\nabla}\!\times\!{\boldsymbol \Lambda}^{-\nu})\cdot ({\boldsymbol {\eta}}_{\nu}({\bf x})+i{\boldsymbol {\zeta}}_{\nu}({\bf x}))\cdot({\boldsymbol\nabla}\!\times\!{\boldsymbol \Lambda}^{\nu})-\frac{1}{2}\!({\boldsymbol\nabla}\!\times\! \dot{{\boldsymbol \Lambda}}^{-\nu})\cdot {\boldsymbol {\eta}}_{\nu}^{\prime\prime}({\bf x})\cdot({\boldsymbol\nabla}\!\times\! \dot{{\boldsymbol \Lambda}}^{\nu})\right. \nonumber\\ 
& \left. -\frac{i}{6}\left(({\boldsymbol\nabla}\!\times\!{\dddot{\boldsymbol \Lambda}}^{-\nu})\cdot {\boldsymbol {\eta}}_{\nu}^{\prime\prime\prime}({\bf x})\cdot({\boldsymbol\nabla}\!\times\!{\boldsymbol \Lambda}^{\nu}) -({\boldsymbol\nabla}\!\times\!{\boldsymbol \Lambda}^{-\nu})\cdot {\boldsymbol {\eta}}_{\nu}^{\prime\prime\prime}({\bf x})\cdot({\boldsymbol\nabla}\!\times\!{\dddot{\boldsymbol \Lambda}}^{\nu})\right)+\mu {\dot{\boldsymbol \Lambda}}^{-\nu}\cdot\dot{\boldsymbol \Lambda}^{\nu}\right]d^{3}{\bf x} \nonumber\\
&+\sum_{n>1}\sum_{\nu, \nu_{1},\cdots,\nu_{n}}\frac{1}{n+1}\int ({\boldsymbol\nabla}\times{\boldsymbol \Lambda}^{\nu}({\bf x},t))\cdot{\boldsymbol {\eta}}^{(n)}({\bf x})\vdots\left(({\boldsymbol\nabla}\times{\boldsymbol \Lambda}^{\nu_1}({\bf x},t))\cdots({\boldsymbol\nabla}\times{\boldsymbol \Lambda}^{\nu_n}({\bf x},t))\right)d^{3}{\bf x}.
\end{align}
By neglecting the third order dispersion term, ${\boldsymbol {\eta}}_{\nu}^{\prime\prime\prime}({\bf x})$ , for a lossless dielectric, the linear Lagragian agrees with Eq.(3.117) in ref. \cite{15}. \\
In summary, the results obtained implementing the total Lagrangian, $L_{T}$, agree in both dynamics and energy with the results obtained from Maxwell's equations and Poynting's theorem for slowly varying envelope functions. So, the total Lagrangian, describing the field propagation through a medium with a combination of dispersion, attenuation, nonlinearity and inhomogeneity, is unique as one can derive the correct equation of motion and the Hamiltonian. Additionally, the linear Lagrangian \ref{eq18} describes the system in the framework of a local field theory of a linear dispersive medium in presence of attenuation. The first and the last terms of the linear Lagrangian and the linear Hamiltonian resemble a massless boson, while the remaining terms indicate dispersive and attenuating corrections.
\section{Field Quantization}
\begin{figure}[h] 
\centering
\includegraphics[width=0.7\columnwidth]{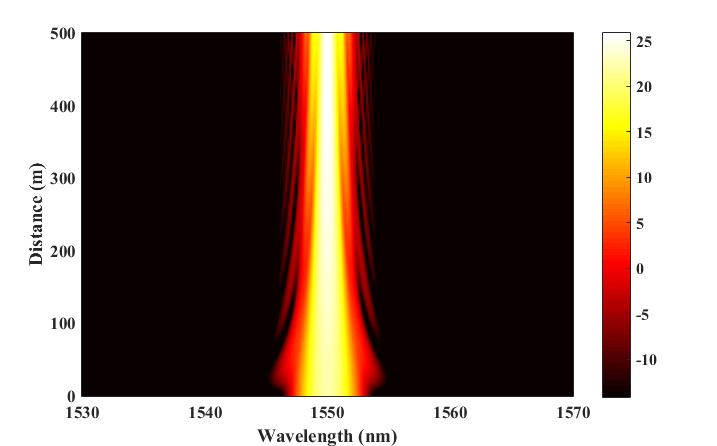}
\caption{Pulse evolution along an slab waveguide of total length $500 m$, when a pulse (central wavelength at $1550 nm$) is propagating along the slab. The attenuation parameter (loss), the nonlinear parameter, the second and the third order dispersion coefficients of the slab at $1550 nm$ are $0.2 dB/km$, $2 W/km$, $-20 ps^{2}/km$ and $0.02 ps^{3}/km$, respectively.} 
\label{fig1}
\end{figure}

\begin{figure}[h] 
\centering
\includegraphics[width=0.7\columnwidth]{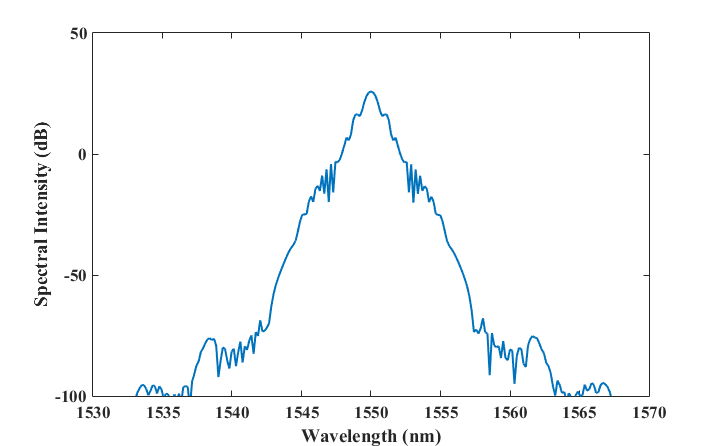}
\caption{Wavelength profile of the pulse at the output of the symmetric one dimensional slab.} 
\label{fig2}
\end{figure}

\begin{figure}[h] 
\centering
\includegraphics[width=0.7\columnwidth]{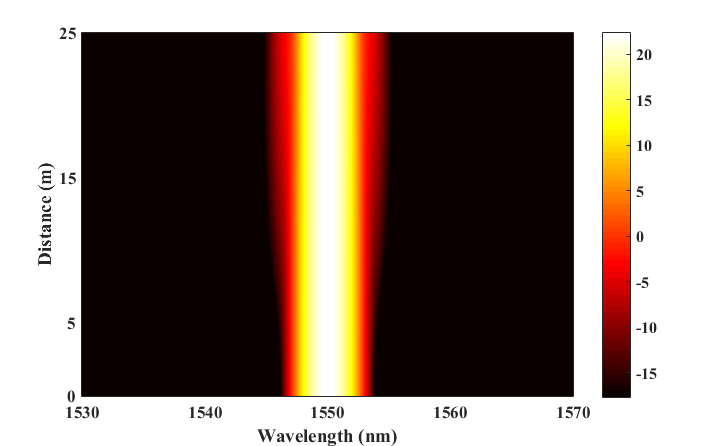}
\caption{Same as fig. \ref{fig1}, but for the first $25 m$ of the slab.} 
\label{fig3}
\end{figure}

According to quantum field theory, quantization procedure is done by  imposing Dirac's commutation relations. For the considered system, Dirac's commutation relations for the components of vector operators $\hat{\boldsymbol{\Lambda}}^{\nu}$ and ${\hat{\boldsymbol{\Pi}}}^{\nu}$ are: 
\begin{equation}  \label{eq23}
\left[{\hat{\Lambda}}_{j}^{\nu}(t,{\bf x}),{\hat{\Pi}}_{j^\prime}^{\nu}(t,{\bf x}^{\prime}) \right]=i\hbar \delta_{jj^{\prime}}\delta^{\perp}({\bf x}-{\bf x}^{\prime}).
\end{equation}
Since the dual potentials and their cononical momenta are transverse, the commutation relations \ref{eq23} are transverse. Equation \ref{eq23} expresses that our system is a kind of constrained system \cite{52} because there is no standard commutation relations. To extend the common approach of quantization to this constrained quantization, it is necessary to construct the appropriate form of the Dirac commutation relations for new coordinates. For this perpose, the dual potential functions are expanded in terms of spatial modes as:
\begin{equation}   \label{eq24}
{{\boldsymbol\Lambda}}^{\nu}(t,{\bf x})=\frac{1}{\sqrt{V}}\sum_{{\bf k},\alpha}\lambda_{{\bf k},\alpha}^{\nu}(t){\hat{\bf e}}_{{\bf k},\alpha}e^{i{\bf k}\cdot{\bf x}}
\end{equation}
to rephrase the constraint, where the expansion coefficients are the new coordinates, $\lambda_{{\bf k},\alpha}^{\nu}$. By inserting the expansion \ref{eq24} into equation \ref{eq18}, the linear part of Lagrangian is obtained as:
\begin{align}  \label{eq25}
L_{L} &=\frac{1}{2}\sum_{\nu=-N}^{+N}\sum_{{\bf k},\alpha}\sum_{{\bf k}^{\prime},\alpha^{\prime}}\left[-(\lambda_{{\bf k}^{\prime},\alpha^{\prime}}^{\nu})^{\ast}    \left(M_{({\bf k}^{\prime},\alpha^{\prime}),({\bf k},\alpha)}^{(1)\nu}+iM_{({\bf k}^{\prime},\alpha^{\prime}),({\bf k},\alpha)}^{(1^\prime)\nu} \right)\lambda_{{\bf k},\alpha}^{\nu} \right. \nonumber \\
&\left. -i\left((\lambda_{{\bf k}^{\prime},\alpha^{\prime}}^{\nu})^{\ast}M_{({\bf k}^{\prime},\alpha^{\prime}),({\bf k},\alpha)}^{(2)\nu}{\dot \lambda}_{{\bf k},\alpha}^{\nu}-(\dot{\lambda}_{{\bf k}^{\prime},\alpha^{\prime}}^{\nu})^{\ast}M_{({\bf k}^{\prime},\alpha^{\prime}),({\bf k},\alpha)}^{(2)\nu}\lambda_{{\bf k},\alpha}^{\nu}\right)\right. \nonumber\\
& \left. +(\dot{\lambda}_{{\bf k}^{\prime},\alpha^{\prime}}^{\nu})^{\ast}M_{({\bf k}^{\prime},\alpha^{\prime}),({\bf k},\alpha)}^{(3)\nu}\dot{\lambda}_{{\bf k},\alpha}^{\nu}+i\left((\ddot{\lambda}_{{\bf k}^{\prime},\alpha^{\prime}}^{\nu})^{\ast}M_{({\bf k}^{\prime},\alpha^{\prime}),({\bf k},\alpha)}^{(4)\nu}{\dot \lambda}_{{\bf k},\alpha}^{\nu}-(\dot{\lambda}_{{\bf k}^{\prime},\alpha^{\prime}}^{\nu})^{\ast}M_{({\bf k}^{\prime},\alpha^{\prime}),({\bf k},\alpha)}^{(4)\nu}\ddot{\lambda}_{{\bf k},\alpha}^{\nu}\right) \right. \nonumber \\
& \left. -\frac{i}{2}\left(({\lambda}_{{\bf k}^{\prime},\alpha^{\prime}}^{\nu})^{\ast}M_{({\bf k}^{\prime},\alpha^{\prime}),({\bf k},\alpha)}^{(4)\nu}{\dddot \lambda}_{{\bf k},\alpha}^{\nu}-(\dddot{\lambda}_{{\bf k}^{\prime},\alpha^{\prime}}^{\nu})^{\ast}M_{({\bf k}^{\prime},\alpha^{\prime}),({\bf k},\alpha)}^{(4)\nu}{\lambda}_{{\bf k},\alpha}^{\nu}\right)\right]
\end{align}
where
\begin{subequations}  \label{eq26}
\begin{align}  \label{eq26a}
M_{({\bf k}^{\prime},\alpha^{\prime}),({\bf k},\alpha)}^{(1)\nu}& =\frac{1}{V}\int({\bf k}^{\prime}\times{\hat{\bf e}}_{{\bf k}^{\prime},\alpha^{\prime}}^{\ast})\cdot {\boldsymbol {\eta}}_{\nu}({\bf x})\cdot ({\bf k}\times{\hat{\bf e}}_{{\bf k},\alpha})e^{i({\bf k}-{\bf k}^\prime)\cdot{\bf x}}d^{3}{\bf x},
\end{align}
\begin{align}  \label{eq26b}
M_{({\bf k}^{\prime},\alpha^{\prime}),({\bf k},\alpha)}^{(1^{\prime})\nu}& =\frac{1}{V}\int({\bf k}^{\prime}\times{\hat{\bf e}}_{{\bf k}^{\prime},\alpha^{\prime}}^{\ast})\cdot {\boldsymbol {\zeta}}_{\nu}({\bf x})\cdot ({\bf k}\times{\hat{\bf e}}_{{\bf k},\alpha})e^{i({\bf k}-{\bf k}^\prime)\cdot{\bf x}}d^{3}{\bf x},
\end{align}
\begin{align}  \label{eq26c}
M_{({\bf k}^{\prime},\alpha^{\prime}),({\bf k},\alpha)}^{(2)\nu}& =\frac{1}{2V}\int({\bf k}^{\prime}\times{\hat{\bf e}}_{{\bf k}^{\prime},\alpha^{\prime}}^{\ast})\cdot {\boldsymbol {\eta}}_{\nu}^{\prime}({\bf x})\cdot ({\bf k}\times{\hat{\bf e}}_{{\bf k},\alpha})e^{i({\bf k}-{\bf k}^\prime)\cdot{\bf x}}d^{3}{\bf x},
\end{align}
\begin{align}  \label{eq26d}
M_{({\bf k}^{\prime},\alpha^{\prime}),({\bf k},\alpha)}^{(3)\nu}& =\left[\mu({\hat{\bf e}}_{{\bf k}^{\prime},\alpha^{\prime}}^{\ast}\cdot{\hat{\bf e}}_{{\bf k},\alpha})\delta_{{\bf k},{\bf k}^{\prime}}-\frac{1}{2V}\int({\bf k}^{\prime}\times{\hat{\bf e}}_{{\bf k}^{\prime\prime},\alpha^{\prime}}^{\ast})\cdot {\boldsymbol {\eta}}_{\nu}^{\prime\prime}({\bf x})\cdot ({\bf k}\times{\hat{\bf e}}_{{\bf k},\alpha})e^{i({\bf k}-{\bf k}^\prime)\cdot{\bf x}}d^{3}{\bf x}\right]
\end{align}
and
\begin{align} \label{eq26e}
M_{({\bf k}^{\prime},\alpha^{\prime}),({\bf k},\alpha)}^{(4)\nu}& =\frac{1}{3V}\int({\bf k}^{\prime}\times{\hat{\bf e}}_{{\bf k}^{\prime},\alpha^{\prime}}^{\ast})\cdot {\boldsymbol {\eta}}_{\nu}^{\prime\prime\prime}({\bf x})\cdot ({\bf k}\times{\hat{\bf e}}_{{\bf k},\alpha})e^{i({\bf k}-{\bf k}^\prime)\cdot{\bf x}}d^{3}{\bf x}.
\end{align}
\end{subequations}
One can simplify the linear Lagrangian \ref{eq25} as (for details see suplementary):
\begin{align}  \label{eq27}
L_{L} &=\sum_{\nu=0}^{+N}\left[-(\lambda^{\nu})^{\dagger}\left(M^{(1)\nu}+iM^{(1^{\prime})\nu}\right)\lambda^{\nu}-i\left((\lambda^{\nu})^{\dagger}M^{(2)\nu}{\dot \lambda}^{\nu}-(\dot{\lambda}^{\nu})^{\dagger}M^{(2)\nu}\lambda^{\nu}\right)+(\dot{\lambda}^{\nu})^{\dagger}M^{(3)\nu}\dot{\lambda}^{\nu}\right. \nonumber\\
& \left. +i\left((\ddot{\lambda}^{\nu})^{\dagger}M^{(4)\nu}{\dot \lambda}^{\nu}-(\dot{\lambda}^{\nu})^{\dagger}M^{(4)\nu}\ddot{\lambda}^{\nu}\right) -\frac{i}{2}\left(({\lambda}^{\nu})^{\dagger}M^{(4)\nu}{\dddot \lambda}^{\nu}-(\dddot{\lambda}^{\nu})^{\dagger}M^{(4)\nu}{\lambda}^{\nu} \right) \right]
\end{align}
by omitting the summations over $({\bf k},\alpha,{\bf k}^{\prime},\alpha^{\prime})$ and the corresponding indices for simplicity.
In order to rephrase the constraint and obtain standard commutation relations, the new canonical momenta corresponding to the new set of coordinates are derived as:
\begin{subequations}    \label{eq28}
\begin{eqnarray}   \label{eq28a}
{\pi}^{\nu}=\frac{\partial L_{L}}{\partial \dot{\lambda}^{\nu}}=\left[-i(\lambda^{\nu})^{\dagger}M^{(2)\nu}+(\dot{\lambda}^{\nu})^{\dagger}M^{(3)\nu}+i(\ddot{\lambda}^{\nu})^{\dagger}M^{(4)\nu}\right]
\end{eqnarray}
and
\begin{eqnarray} \label{eq28b}
({\pi}^{\nu})^{\dagger}=\frac{\partial L_{L}}{\partial (\dot{\lambda}^{\nu})^{\dagger}}=\left[iM^{(2)\nu}\lambda^{\nu}+M^{(3)\nu}\dot{\lambda}^{\nu}-iM^{(4)\nu}\ddot{\lambda}^{\nu}\right].
\end{eqnarray}
\end{subequations}
It is straightforward to find the linear part of Hamiltonian in terms of the new canonical coordinates and momenta as:
\begin{align} \label{eq29}
H_{L}&=\sum_{\nu=0}^{+N}\left\lbrace \!\left[\! \left(\!\pi^{\nu}\!+\! i(\!\lambda^{\nu}\!)^{\dagger}\!M^{(2)\nu}\right)\!\left(\!M^{(3)\nu}\! \right)^{-1}\!\!\left(\!(\!\pi^{\nu}\!)^{\dagger}\!-\!iM^{(2)\nu}\lambda^{\nu} \right)\!+\!(\!\lambda^{\nu}\!)^{\dagger}\left(\!M^{(1)\nu}\!+\!iM^{(1^{\prime})\nu}  \right)\!\lambda^{\nu}\!\right] \right. \nonumber\\
&\left.+\!\left(\!\pi^{\nu}\!+\!i(\!\lambda^{\nu}\!)^{\dagger}\!M^{(2)\nu}\right)\!\left(\!M^{(3)\nu}\!\right)^{-1}\!\left(\!iM^{(4)\nu}\ddot{\lambda}^{\nu} \!\right)\!-\!\left(\!i(\!\ddot{\lambda}^{\nu}\!)^{\dagger}M^{(4)\nu}\!\right)\!\left(M^{(3)\nu}\!\right)^{-1}\!\left(\!(\!\pi^{\nu}\!)^{\dagger}\!-\!iM^{(2)\nu}\!\lambda^{\nu}\! \right)\! \right. \nonumber\\
&\left.+\!(\!\ddot{\lambda}^{\nu}\!)^{\dagger}M^{(4)\nu}\left(\!M^{(3)\nu} \!\right)^{-1}M^{(4)\nu} \ddot{\lambda}^{\nu}\!+\!\frac{i}{2}\left(\!(\!\lambda^{\nu})^{\dagger}M^{(4)\nu}\dddot{\lambda}^{\nu}\!-\!(\!\dddot{\lambda}^{\nu}\!)^{\dagger}M^{(4)\nu}{\lambda}^{\nu}\!\right)\!\right\rbrace.
\end{align}
It is practical to rewrite $H_{L}$ as:
\begin{equation}  \label{eq30}
H_{L}=H_{Ld}+H_{La}
\end{equation}
 where
 \begin{subequations}       \label{eq31}
 \begin{align} \label{eq31a}
H_{Ld}&=\sum_{\nu=0}^{+N}\left\lbrace \!\left[\! \left(\!\pi^{\nu}\!+\! i(\!\lambda^{\nu}\!)^{\dagger}\!M^{(2)\nu}\!\right)\!\left(\!M^{(3)\nu}\! \right)^{-1}\!\!\left(\!(\!\pi^{\nu}\!)^{\dagger}\!-\!iM^{(2)\nu}\!\lambda^{\nu} \right)\!+\!(\!\lambda^{\nu}\!)^{\dagger}M^{(1)\nu}\!\lambda^{\nu}\!\right] \right. \nonumber\\
&\left.+\!\left(\!\pi^{\nu}\!+\!i(\!\lambda^{\nu}\!)^{\dagger}\!M^{(2)\nu}\!\right)\!\left(\!M^{(3)\nu}\!\right)^{-1}\!\left(\!iM^{(4)\nu}\ddot{\lambda}^{\nu} \!\right)\!-\!\left(\!i(\!\ddot{\lambda}^{\nu}\!)^{\dagger}M^{(4)\nu}\!\right)\!\left(M^{(3)\nu}\!\right)^{-1}\!\left(\!(\!\pi^{\nu}\!)^{\dagger}\!-\!iM^{(2)\nu}\!\lambda^{\nu}\! \right)\! \right. \nonumber\\
&\left.+(\!\ddot{\lambda}^{\nu}\!)^{\dagger}M^{(4)\nu}\left(\!M^{(3)\nu} \!\right)^{-1}M^{(4)\nu} \ddot{\lambda}^{\nu}\!+\!\frac{i}{2}\left(\!(\!\lambda^{\nu})^{\dagger}M^{(4)\nu}\dddot{\lambda}^{\nu}\!-\!(\!\dddot{\lambda}^{\nu}\!)^{\dagger}M^{(4)\nu}{\lambda}^{\nu}\!\right)\!\right\rbrace.
\end{align}
and
\begin{align} \label{eq31b}
H_{La}&=i\sum_{\nu=0}^{+N}(\lambda^{\nu}\!)^{\dagger}M^{(1^{\prime})\nu}\lambda^{\nu}
\end{align}
\end{subequations}
describe dispersion of the dielectric up to third order term and the dielectric's attenuation, respectively. Obtaining  $H_{L}$ is one of the goals of this paper which it is reduced to Eq. (3.132) of \cite{15} when the third order dispersion and the attenuation are neglected.\\
It is necessary to impose the standard commutation relations between $\lambda^{\nu}$ and $\pi^{\nu}$, to quantize the fields for the current problem. These relations no longer have transversality restrictions as compared with the operators $\hat{\boldsymbol \Lambda}^{\nu}$ and $\hat{\boldsymbol \Pi}^{\nu}$. The commutation relations between $\hat{\lambda}^{\nu}$ and $\hat{\pi}^{\nu}$ can be simply written as: 
\begin{equation}  \label{eq32}
\left[\hat{\lambda}_{{\bf k},\alpha}^{\nu},\hat{\pi}_{{\bf k}^\prime,\alpha^\prime}^{\nu} \right]=i\hbar \delta_{{\bf k}{{\bf k}^\prime}}\delta_{\alpha{\alpha^\prime}}.
\end{equation}
Using boson's creation and annihilation operators, it is possible to re-expand the linear Hamiltonian. These operators are defined as column vectors
\begin{equation}  \label{eq33}
{\hat{\bf a}}^{\nu}=\frac{1}{\sqrt{2\hbar}}\left[A^{\nu}\cdot {\hat{\lambda}}^{\nu}+i\left( \left(A^{\nu} \right)^{\dagger} \right)^{-1}\cdot\left({\hat{\pi}}^{\nu} \right)^{\dagger}  \right]
\end{equation}
and
\begin{equation}  \label{eq34}
\left( {\hat{\bf b}}^{\nu}\right) ^{\dagger}=\frac{1}{\sqrt{2\hbar}}\left[A^{\nu}\cdot {\hat{\lambda}}^{\nu}-i\left( \left(A^{\nu} \right)^{\dagger} \right)^{-1}\cdot\left({\hat{\pi}}^{\nu} \right)^{\dagger}\right],
\end{equation}
where the transformation matrix ${A}^{\nu}$ is an arbitrary invertible complex matrix to be defined. On the basis of quantum field theory, it is found that the operators ${\hat{\bf{a}}}^{\nu}$ and ${\hat{\bf{b}}}^{\nu}$ have the character of annihilation operator, while $\left({\hat{\bf{a}}}^{\nu}\right)^{\dagger}$ and $\left({\hat{\bf{b}}}^{\nu}\right)^{\dagger}$ are creation operators. Commutation relations for these two operators are: 
\begin{eqnarray}  \label{eq35}
\left[{\hat{\bf a}}_{i}^{\nu},\left( {\hat{\bf a}}_{j}^{\nu}\right) ^{\dagger}\right] =\left[{\hat{\bf b}}_{i}^{\nu},\left( {\hat{\bf b}}_{j}^{\nu}\right) ^{\dagger}\right] =\delta_{ij}
\end{eqnarray}
and
\begin{eqnarray}  \label{eq36}
\left[{\hat{\bf a}}_{i}^{\nu},{\hat{\bf a}}_{j}^{\nu} \right]=\left[{\hat{\bf b}}_{i}^{\nu},{\hat{\bf b}}_{j}^{\nu}\right]=\left[\left( {\hat{\bf a}}_{i}^{\nu}\right) ^{\dagger},{\hat{\bf b}}_{j}^{\nu} \right]=\left[{\hat{\bf a}}_{i}^{\nu}, \left( {\hat{\bf b}}_{j}^{\nu}\right) ^{\dagger}\right]=0.
\end{eqnarray}
For the case $\nu=0$, one can conclude that ${\hat{\bf{a}}}^{0}$ and ${\hat{\bf{b}}}^{0}$ are not independent operators when the Hamiltonian \ref{eq29} describes a lossless nondispersive medium, therefore, only ${\hat{\bf{a}}}^{0}$ will be used. The Hamiltonian \ref{eq29} can be written  in terms of the creation and the annihilation operators as:
\begin{equation}  \label{eq37}
{\hat{H}}_{L}= \hbar \sum_{\nu=0}\left( {\hat{\bf a}}^{\nu}\right) ^{\dagger}\cdot{\Omega^{\nu}}\cdot{\hat{\bf a}}^{\nu}+\hbar \sum_{\nu=1}\left( {\hat{\bf b}}^{\nu}\right) ^{\dagger}\cdot{\Omega^{-\nu}}\cdot{\hat{\bf b}}^{\nu},
\end{equation}
where $\Omega^{\pm\nu}$ are defined as frequency matrices and the relations:
\begin{subequations}     \label{eq38}
\begin{align}  \label{eq38a}
\frac{1}{2}\left(A^{\nu}\right)^{-1}\cdot \left( \!\Omega^{+\nu}\!+\Omega^{-\nu}\!\right)\cdot\left(\left(A^{\nu}\right)^{\dagger}\right)^{-1}&\!=\!\left(M^{(3)} \right)^{-1},
\end{align}
\begin{align} \label{eq38b}
\frac{1}{2}\left(A^{\nu}\right)^{\dagger}\cdot \left(\! \Omega^{+\nu}\!-\!\Omega^{-\nu}\!\right)\cdot\left(\left(A^{\nu}\right)^{\dagger}\right)^{-1}&\!=\!\!\left[\!M^{(2)\nu}\!+\!\left(\Omega^{-\nu}\right)^{2}\cdot M^{(4)\nu}\!\right] \cdot \left(M^{(3)\nu}\right)^{-1}
\end{align}
and
\begin{align} \label{eq38c}
\frac{1}{2}\left(A^{\nu}\right)^{\dagger}\cdot \left(\!\Omega^{+\nu}\!+\!\Omega^{-\nu}\!\right)\cdot A^{\nu}&=\!\left[\!M^{(1)\nu}\!+\!iM^{(1^{\prime})\nu}\!+\!M^{(2)\nu}\!\cdot \!\left(\!M^{(3)\nu}\!\right)^{\!-1}\!\cdot\! M^{(2)\nu}\! +\!M^{(2)\nu}\!\cdot\! \left(\!M^{(3)\nu}\!\right)^{\!-1}\!\cdot\! M^{(4)\nu}\!\cdot\! \left(\Omega^{+\nu}\right)^{2}\! \right.  \nonumber \\
& \left. +\! \left(\Omega^{-\nu}\right)^{2}\!\cdot\! M^{(4)\nu}\!\cdot\! \left(\!M^{(3)\nu}\!\right)^{\!-1}\!\cdot\! M^{(2)\nu}\!+\!\left(\Omega^{-\nu}\right)^{2}\!\cdot\! M^{(4)\nu}\cdot \left(\!M^{(3)\nu}\!\right)^{-1}\!\cdot\! M^{(4)\nu}\!\cdot\! \left(\Omega^{+\nu}\right)^{2}\! \right. \nonumber \\
&\left. -\frac{1}{2}\!\left(M^{(4)\nu}\!\cdot\!\left(\Omega^{+\nu}\right)^{3}-\left(\Omega^{-\nu}\right)^{3}\!\cdot\! M^{(4)\nu}\right)\!\right]=F^{\nu} 
\end{align}
\end{subequations}
hold by equating two forms of Hamiltonian \ref{eq29} and \ref{eq37}, while neglecting zero point energy. To calculate the equations \ref{eq38}, the Heisenberg's equation and the Hamiltonian \ref{eq37} are applied for obtaining the dynamics of the operators as ${\hat{\dot{\bf a}}}^{\nu}=-i\Omega^{+\nu}\cdot {\hat{\bf a}}^{\nu}$ (or equivalently ${\hat{\dot{\lambda}}}^{\nu}=-i\Omega^{+\nu}\cdot {\hat{\lambda}}^{\nu}$) and ${\hat{\dot{\bf b}}}^{\nu}=-i{\hat{\bf b}}^{\nu}\cdot \Omega^{-\nu}$ (or equivalently $({\hat{\dot{\lambda}}}^{\nu})^{\dagger}=-i({\hat{\lambda}}^{\nu})^{\dagger}\cdot \Omega^{-\nu}$). So, it is understood that $(\Omega^{+\nu})^{\dagger}=-\Omega^{-\nu}$.\\
The equations \ref{eq38} are valid, if the determinants of the matrices $M^{(i)\nu}$, $\Omega^{\pm\nu}$ and $A^{\nu}$ are non-zero. The frequency matrices $\Omega^{+\nu}$ and $\Omega^{-\nu}$ should have no zero eigenvalues for this theory to be valid, otherwise, one needs to change the formalism to exclude the zero eigenvalues of these matrices.
The relation: 
\begin{align}    \label{eq39}
\left[{\left(\!A^{\nu}\!\right)}^{\dagger}\!\!\cdot\! A^{\nu}\!\cdot\!\left(\!M^{(3)\nu}\!\right)^{-\!1}\right]^{2}&\!= F^{\nu}\!\!\cdot\!\left(\!M^{(3)\nu}\!\right)^{-\!1}
\end{align}
holds for the matrices $A^{\nu}$ by eliminating $\Omega^{\pm\nu}$ in equations \ref{eq38}. It could be shown that equation \ref{eq37} holds only when $A^{\nu}$ is a solution to quartic matrix equation \ref{eq39}. The corresponding frequency matrices are found as:
\begin{equation}   \label{eq40}
\Omega^{\pm\nu}=\left[A^{\nu}\!\pm\! \left({\left(A^{\nu}\right)}^{\dagger}\right)^{-1}\cdot \left[
M^{(2)\nu}+\!\left(\Omega^{-\nu}\right)^{2}M^{(4)\nu}\right]\right] \cdot\left(M^{(3)\nu}\right)^{-1}\cdot \left(A^{\nu}\right)^{\dagger}.
\end{equation}
In general, the resulted Hamiltonian is not diagonal as $M^{(1)\nu}$ to $M^{(4)\nu}$ matrices are not diagonal. The matrices $\Omega^{\pm\nu}$ could be diagonalized to obtain different frequency bands as: 
\begin{equation} \label{eq41}
\left[U^{\nu}\cdot\Omega^{\pm \nu}\cdot\left(U^{\nu}\right)^{-1}\right]_{nm}=\omega_{n}^{\pm \nu}\delta_{nm}
\end{equation}
resulting in the diagonalized Hamiltonian: 
\begin{equation}     \label{eq42} \hat{H_{L}}=\hbar\sum_{\nu=0}^{N}\sum_{n}\omega_{n}^{\nu}\left({{\tilde{{\bf a}}}_{n}^{\nu}}\right)^{\dagger}\cdot{{\tilde{{\bf a}}}_{n}^{\nu}}+\hbar\sum_{\nu=1}^{N}\sum_{n}\omega_{n}^{-\nu}({{\tilde{{\bf b}}}_{n}^{\nu}})^{\dagger}\cdot{{\tilde{{\bf b}}}_{n}^{\nu}}
\end{equation}
where
\begin{equation}  \label{eq43}
{{\tilde{{\bf a}}}_{n}^{\nu}}=U^{\nu}\cdot{{\hat{{\bf a}}}_{n}^{\nu}}
\end{equation}
and
\begin{equation}  \label{eq44}
{{\tilde{{\bf b}}}_{n}^{\nu}}=U^{\nu}\cdot{{\hat{{\bf b}}}_{n}^{\nu}}.
\end{equation}
Implementing this method, two sets of modal solutions for the diagonalized operators ${\tilde{\bf{a}}}^{\nu}$ and ${\tilde{\bf{b}}}^{\nu}$ are resulted. A set of normal modes corresponds to the diagonal operators ${\tilde{\bf{a}}}^{\nu}$, and a set of anomalous modes corresponds to the diagonal operators ${\tilde{\bf{b}}}^{\nu}$. The anomalous modes generate envelopes comprising of negative frequencies while the Taylor's expansion will not be valid. However, these anomalous modes are a part of the Lagrangian. Therefore, as their corresponding dynamics are outside the range of validity of the used approximations, so we neglect them. In equation \ref{eq42}, the operators indicate the numbers of photon-polaritons of the system. It should be noted that the quantities $M^{(1)\nu}$ to $M^{(3)\nu}$ defined by equations \ref{eq26}, change when the third order dispersion and the attenuation are absent while $M^{(4)\nu}$ and $M^{(1^\prime)\nu}$ are zero.\\
By neglecting the anomalous modes, one can calculate the nonlinear Hamiltonian,
\begin{align} \label{eq45} 
\hat{H}_{NL}&=\frac{1}{4}\int{\hat{{\bf D}}}(t, {\bf x})\cdot\eta^{(3)}({\bf x})\vdots{\hat{{\bf D}}}(t, {\bf x}){\hat{{\bf D}}}(t, {\bf x}){\hat{{\bf D}}}(t, {\bf x})d^{3}{\bf x},
\end{align}
making use of the dual potentials in terms of annihilation operators 
\begin{equation}  \label{eq46}
\hat{\boldsymbol{\Lambda}}^{\nu}(t,{\bf x})=\sqrt{\frac{\hbar}{2V}}\sum_{{\bf k}}\sum_{{\bf k}^{\prime}}{\hat{\bf e}}_{{\bf k}}e^{i{\bf k}\cdot{\bf x}}\left(A^{\nu}\right)_{{\bf k}{\bf k}^{\prime}}^{-1}{\hat{\bf a}}_{{\bf k}}^{\nu}(t)
\end{equation}
where ${\hat{\boldsymbol \Lambda}}^{-\nu}(t, {\bf x})=\left({\hat{\boldsymbol \Lambda}}^{\nu}(t, {\bf x})\right)^{\dagger}$. Therefore, the displacement and the magnetic field operators in terms of the annihilation and the creation operators are:
\begin{subequations}       \label{eq47}
\begin{equation}  \label{eq47a}
\hat{\bf D}(t,{\bf x})=\sqrt{\frac{\hbar}{2V}}\sum_{{\nu=0}}\left(\sum_{{{\bf k}}}\sum_{{{\bf k}}^{\prime}}\left({\bf k}\!\times\!{\hat{\bf e}}_{{\bf k}}\right)e^{i{\bf k}\cdot{\bf x}}\left(A^{\nu}\right)_{{\bf k}{\bf k}^{\prime}}^{-1}{\hat{\bf a}}_{{\bf k}}^{\nu}(t)-h.c.\right)
\end{equation}
and
\begin{equation}  \label{eq47b}
\hat{\bf B}(t,{\bf x})=\mu \hat{\dot{\boldsymbol \Lambda}}(t, {\bf x}),
\end{equation}
\end{subequations}
respectively. The established theory could properly quantize the electromagnetic radiation in a three dimensional dielectric in presence of third order
dispersion and attenuation term.\\
By inserting Eq. \ref{eq47a} into Eq. \ref{eq45}, the nonlinear Hamiltonian will be obtained in terms of the annihilation and creation operators. To calculate the nonlinear Hamiltonian correctly, it is necessary to know the detailed information on the nonlinear parameters of the medium.\\
As $F^{\nu}$ is a complex quantity due to presence of the attenuation term, $M^{(1^{\prime})\nu}$, $A^{\nu}$ and $\omega_{n}^{ \nu}$ are complex, consequently. By neglecting the anomalous modes, the Hamiltonian \ref{eq42} can be re-written as:
\begin{equation} \label{eq48} \hat{H_{L}}=\hbar\sum_{\nu=0}^{N}\sum_{n}\gamma_{n}^{\nu}\left({{\tilde{{\bf a}}}_{n}^{\nu}}\right)^{\dagger}\cdot{{\tilde{{\bf a}}}_{n}^{\nu}}+i\hbar\sum_{\nu=0}^{N}\sum_{n}\alpha_{n}^{\nu}\left({{\tilde{{\bf a}}}_{n}^{\nu}}\right)^{\dagger}\cdot{{\tilde{{\bf a}}}_{n}^{\nu}}
\end{equation}
where $\gamma_{n}^{\nu}=\Re(\omega_{n}^{\nu})$ and $\alpha_{n}^{\nu}=\Im(\omega_{n}^{\nu})$. Applying the method introduced in \cite{42}, one can obtain a coupled stochastic nonlinear Schr\"{o}dinger equations describing pulse propagation through a one dimensional slab quantum mechanically in the presence of the attenuation term (loss). A Gaussian pulse of peak power $10 W$ and the width of $1 p\!s$ is assumed to be launched into a slab waveguide and the evolution of the pulse along the slab is simulated in the mean case \cite{53} as shown in fig. \ref{fig1}. The wavelength dependence of the pulse exiting the slab of $500 m$ long is shown in fig. \ref{fig2}, while fig. \ref{fig3} presents the evolution of the pulse in the first $25 m$ of the slab. See the detailed characteristics of the slab in the caption of fig. \ref{fig1}. 
\section{Concluding Remarks}
In summary, a canonical Lagrangian was defined to include the third order dispersion and the attenuation coefficients by introducing a dual potential. This Lagrangian resulted in the correct equation of motion and the classical energy for the field propagation through an anisotropic inhomogeneous dispersive dielectric where the third order nonlinear term is not negligible. Using the derived Lagrangian, a new set of canonical coordinates and momenta were obtained while commutation relations were derived employing constraint quantization approach. In turn, the creation and the annihilation operators were defined obeying the proper commutation relations. According to the presented theory, we acquired the total Hamiltonian for the field propagation through anisotropic inhomogeneous dispersive dielectric and quantized the electric displacement and magnetic fields. Hence, by defining proper creation and annihilation operators for the linear and nonlinear parts of the system, we derived the Hamiltonian in terms of these operators. This enabled us to include the third order dispersion and the attenuation into field propagation in dielectrics. This theory can be applied to the propagation of quantum solitons and their interactions in the presence of the third order dispersion and the attenuation terms by using the method presented in \cite{42}. Additionally, in resemblance to other physical phenomena, there are noises present in nonlinear optical phenomena including quantum noise. This work could be applied to reduce these fluctuations and quantum noise in the vicinity of the solitons formed in the presence of the third order dispersion and the attenuation coefficients in dielectrics such as fibers \cite{42}. Furthermore, the supercontinuum generation as well as soliton self-frequency shift phenomena could be studied quantum mechanically, if the retarded nonlinear response of the medium is considered. 

\section*{Appendix}
\setcounter{equation}{0}
\renewcommand{\theequation}{A.\arabic{equation}}
Here, we present a one-dimensional example of the problem for a plane wave propagating along the z-direction and containing a single transverse mode $\nu_{0}$. The dual field is defined as, $\boldsymbol{\Lambda}(t,\textbf{x})=\Lambda(t,z)\hat{x}$, where the electric displacement vector, the magnetic field and the integration over the quantization volume are reduced to, ${\bf D}(t,{\bf x})=\hat{y}\partial_{z}{\Lambda}(t,z)$, ${\bf B}(t,{\bf x})=\mu \dot{\boldsymbol{\Lambda}}(t,{\bf x})=\hat{x}\mu\partial_{t}{\Lambda}(t,z)$ and $A\int_{0}^{L}\cdots dz$, respectively, for the transverse area, $A$. It is assumed that the response tensors of the medium are homogenous, isotropic and the first non-zero nonlinear term corresponds to $\eta^{(3)}$ (centro-symmetric media). Therefore, the total Hamiltonian for the field propagation is simplified as:
\begin{align}     \label{eq-A-1}
H&=\frac{A}{2}\int dz\left\lbrace \! \mu\dot{\Lambda}^{-\nu_{0}}\dot{\Lambda}^{\nu_{0}}\!+\!(\eta_{\nu_{0}}+i\zeta_{\nu_{0}})\partial_{z}{\Lambda}^{-\nu_{0}}\partial_{z}{\Lambda}^{\nu_{0}}\!-\!\frac{1}{2}\eta^{\prime\prime}_{\nu_{0}}\partial_{z}\dot{\Lambda}^{-\nu_{0}}\partial_{z}\dot{\Lambda}^{\nu_{0}} \right. \nonumber \\
& \left.-\!\frac{i}{6}\!\eta^{\prime\prime\prime}_{\nu_{0}}\left(\partial_{z}{\Lambda}^{-\nu_{0}}\partial_{z}{\dddot{\Lambda}}^{\nu_{0}}\!-\!\partial_{z}{\Lambda}^{\nu_{0}}\partial_{z}{\dddot{\Lambda}}^{-\nu_{0}}\right)\!+\!\frac{\eta^{(3)}}{2}\!\!\left[\partial_{z}\left(\Lambda^{\nu_{0}}\!+\!\Lambda^{-\nu_{0}}\right)\right]^{4} \! \right\rbrace 
\end{align}
The scaler field $\Lambda(t,z)$ for longitudinal modes $k$ is given as:
\begin{equation}   \label{eq-A-2}
\Lambda(t,z)=\frac{1}{\sqrt{L}}\sum_{k}\left(\lambda_{k}^{\nu_{0}}e^{ikz}+\lambda_{-k}^{-\nu_{0}}e^{-ikz}\right) 
\end{equation}
in terms of the new canonical coordinates, $\lambda_{k}^{\nu_{0}}$. Writing equations \ref{eq27} and \ref{eq28} for the present single dimensional case, the Lagrangian is given by:
\begin{align}    \label{eq-A-3}
L_{L}&\!\!=\sum_{k}\left[\!-(\lambda_{k}^{\nu_{0}})^{\dagger}\lambda_{k}^{\nu_{0}}k^{2}(\eta_{\nu_{0}}+i\zeta_{\nu_{0}}) -\!\frac{i}{2}\!\left((\lambda_{k}^{\nu_{0}})^{\dagger}{\dot{\lambda}}_{k}^{\nu_{0}}\!-\!({\dot{\lambda}}_{k}^{\nu_{0}})^{\dagger}\lambda_{k}^{\nu_{0}}\right)\!k^{2}\eta^{\prime}_{\nu_{0}}\!+\!({\dot{\lambda}}_{k}^{\nu_{0}})^{\dagger}{\dot{\lambda}}_{k}^{\nu_{0}}\left(\mu\!-\!\frac{1}{2}k^{2}\eta^{\prime\prime}_{\nu_{0}} \right)\right. \nonumber\\
&\left. +\!\frac{i}{3}\!\left(({\ddot{\lambda}}_{k}^{\nu_{0}})^{\dagger}{\dot{\lambda}}_{k}^{\nu_{0}}\!-\!({\dot{\lambda}}_{k}^{\nu_{0}})^{\dagger}{\ddot{\lambda}}_{k}^{\nu_{0}}\right)\!k^{2}\eta^{\prime\prime\prime}_{\nu_{0}}\!-\!\frac{i}{6}\!\left((\lambda_{k}^{\nu_{0}})^{\dagger}{\dddot{\lambda}}_{k}^{\nu_{0}}\!-\!({\dddot{\lambda}}_{k}^{\nu_{0}})^{\dagger}\lambda_{k}\right)\!k^{2}\eta^{\prime\prime\prime}_{\nu_{0}}\!\right]
\end{align}
where the canonical momenta associated with $\lambda_{k}$ and $\lambda_{k}^{\dagger}$ are 
\begin{subequations}    \label{eq-A-4}
\begin{align}  \label{eq-A-4a}
\pi_{k}^{\nu_{0}}&=\left[-\frac{i}{2}k^{2}\eta^{\prime}_{\nu_{0}}(\lambda_{k}^{\nu_{0}})^{\dagger}+\left(\mu-\frac{1}{2}k^{2}\eta^{\prime\prime}_{\nu_{0}} \right)({\dot{\lambda}}_{k}^{\nu_{0}})^{\dagger}+\frac{i}{3}k^{2}\eta^{\prime\prime\prime}_{\nu_{0}}({\ddot{\lambda}}_{k}^{\nu_{0}})^{\dagger}\right]
\end{align}
and
\begin{align}   \label{eq-A-4b}
(\pi_{k}^{\nu_{0}})^{\dagger}&=\left[\frac{i}{2}k^{2}\eta^{\prime}_{\nu_{0}}\lambda_{k}^{\nu_{0}}+\left(\mu-\frac{1}{2}k^{2}\eta^{\prime\prime}_{\nu_{0}} \right){\dot{\lambda}}_{k}^{\nu_{0}}-\frac{i}{3}k^{2}\eta^{\prime\prime\prime}_{\nu_{0}}{\ddot{\lambda}}_{k}^{\nu_{0}}\right]
\end{align}
\end{subequations}
respectively. As the dual potential and $\dot{\Lambda}$ are Hermitian, the relations $\lambda_{-k}^{-\nu_{0}}=(\lambda_{k}^{\nu_{0}})^{\dagger}$ and $\pi_{-k}^{-\nu_{0}}=(\pi_{k}^{\nu_{0}})^{\dagger}$ must hold for the new canonical coordinates and momenta. Similarly, one can find the Hamiltonian, $H_{L}$, as:
\begin{align}  \label{eq-A-5}
H_{L}&\!\!=\!\sum_{k}\!\left[\!\left(\!\mu\!-\!\frac{1}{2}k^{2}\eta^{\prime\prime}_{\nu_{0}}\!\right)^{-1}\!\!\left(\pi_{k}^{\nu_{0}}\!+ik^{2}\eta^{\prime}_{\nu_{0}}(\lambda_{k}^{\nu_{0}})^{\dagger}\right)\left((\pi_{k}^{\nu_{0}})^{\dagger}\!-\!ik^{2}\eta^{\prime}_{\nu_{0}}\lambda_{k}^{\nu_{0}}\right) +k^{2}(\eta_{\nu_{0}}+i\zeta_{\nu_{0}})(\lambda_{k}^{\nu_{0}})^{\dagger}\lambda_{k}^{\nu_{0}}\!\!\right. \nonumber\\
&\left.+ik^{2}\eta^{\prime\prime\prime}_{\nu_{0}}\left(\!\mu\!-\!\frac{1}{2}k^{2}\eta^{\prime\prime}_{\nu_{0}}\!\right)^{-1}\!\!\left(\pi_{k}^{\nu_{0}}\!+\!ik^{2}\eta^{\prime}_{\nu_{0}}(\lambda_{k}^{\nu_{0}})^{\dagger}\right)\!{\ddot{\lambda}}_{k}^{\nu_{0}}-ik^{2}\eta^{\prime\prime\prime}_{\nu_{0}}\left(\mu-\frac{1}{2}k^{2}\eta^{\prime\prime}_{\nu_{0}}\right)^{-1}\!\!({\ddot{\lambda}}_{k}^{\nu_{0}})^{\dagger}\left((\pi_{k}^{\nu_{0}})^{\dagger}-ik^{2}\eta^{\prime}_{\nu_{0}}\lambda_{k}^{\nu_{0}}\right)\right. \nonumber\\
&\left.+\left(\!\mu\!-\!\frac{1}{2}\!k^{2}\!\eta^{\prime\prime}_{\nu_{0}}\!\right)^{-1}\!\!\!\left(\!k^{2}\!\eta^{\prime\prime\prime}_{\nu_{0}}\!\right)^{2}\!({\ddot{\lambda}}_{k}^{\nu_{0}})^{\dagger}_{k}{\ddot{\lambda}}_{k}^{\nu_{0}}+\frac{i}{2}\!k^{2}\!\eta^{\prime\prime\prime}_{\nu_{0}}\left(\!({\lambda}_{k}^{\nu_{0}})^{\dagger}{\dddot{\lambda}}_{k}^{\nu_{0}}-({\dddot{\lambda}}_{k}^{\nu_{0}})^{\dagger}{{\lambda}}_{k}^{\nu_{0}} \right) \right]  
\end{align}
where the linear part of the field is assumed and the $M$ quantities (equations \ref{eq26}) are simplified to diagonal matrices as: 
\begin{subequations}            \label{eq-A-6}
\begin{align}                \label{eq-A-6a}
M^{(1)\nu_{0}}_{kk^{\prime}}&=k^{2}\eta_{\nu_{0}}\delta_{kk^{\prime}},
\end{align}
\begin{align}                \label{eq-A-6b}
M^{(1^\prime)\nu_{0}}_{kk^{\prime}}&=k^{2}\zeta_{\nu_{0}}\delta_{kk^{\prime}},
\end{align}
\begin{align}    \label{eq-A-6c}
M^{(2)\nu_{0}}_{kk^{\prime}}&=\frac{1}{2}k^{2}\eta^{\prime}_{\nu_{0}}\delta_{kk^{\prime}},
\end{align}
\begin{align}     \label{eq-A-6d}
M^{(3)\nu_{0}}_{kk^{\prime}}&=\left(\mu-\frac{1}{2}k^{2}\eta^{\prime\prime}_{\nu_{0}} \right)\delta_{kk^{\prime}}
\end{align}
and
\begin{align}   \label{eq-A-6e}
M^{(4)\nu_{0}}_{kk^{\prime}}&=\frac{1}{3}k^{2}\eta^{\prime\prime\prime}_{\nu_{0}}\delta_{kk^{\prime}}.
\end{align}
\end{subequations}
In this one dimensional example, the field is quantized when operators ${\hat{\lambda}}_{k}^{\nu_{0}}$ and ${\hat{\pi}}_{k}^{\nu_{0}}$ obey the commutation relation:
\begin{equation}     \label{eq-A-7}
\left[{\hat{\lambda}}_{k}^{\nu_{0}},{\hat{\pi}}^{\nu_{0}}_{k^{\prime}}\right]=i\hbar\delta_{kk^{\prime}}.
\end{equation}
An annihilation operator is also defined as: 
\begin{equation}  \label{eq-A-8}
{\hat{a}}_{k}^{\nu_{0}}=\frac{1}{\sqrt{2\hbar}}\left(A_{k}^{\nu_{0}}{\hat{\lambda}}_{k}^{\nu_{0}}+i\left((A_{k}^{\nu_{0}})^{\star}\right)^{-1}({\hat{\pi}}_{k}^{\nu_{0}})^{\dagger} \right) 
\end{equation}
where $A_{k}^{\nu_{0}}$ is generally a complex number. Likewise the creation operator $({\hat{a}}_{k}^{\nu_{0}})^{\dagger}$ could be defined.
Therefore, the linear Hamiltonian is resulted as:
\begin{equation}   \label{eq-A-9}
{\hat{H}}_{L}=\hbar\sum_{k}\omega(k)({\hat{a}}_{k}^{\nu_{0}})^{\dagger}{\hat{a}}_{k}^{\nu_{0}},
\end{equation}
while $\omega(k)$ is the solution to equations:
\begin{align}   \label{eq-A-10}
\omega(k)&=\vert A_{k}^{\nu_{0}}\vert^{2}\left(\mu-\frac{1}{2}k^{2}\eta^{\prime\prime}_{\nu_{0}}\right)^{-1}+\frac{1}{2}\left[k^{2}\eta^{\prime}_{\nu_{0}}+\frac{2}{3}k^{2}\eta^{\prime\prime\prime}_{\nu_{0}} \omega^{2}\right]\left(\mu-\frac{1}{2}k^{2}\eta^{\prime\prime}_{\nu_{0}}\right)^{-1}
\end{align}
and
\begin{equation}    \label{eq-A-11}
\vert\omega\vert^{2}=k^{2}c^{2}\left(\eta_{\nu_{0}}+\zeta_{\nu_{0}}+\omega \eta^{\prime}_{\nu_{0}}+\frac{1}{2}\omega^{2}\eta^{\prime\prime}_{\nu_{0}}+\frac{1}{6}\omega^{3}\eta^{\prime\prime\prime}_{\nu_{0}} \right) 
\end{equation}
Substituting the $M$ values, equations \ref{eq-A-6}, into equation \ref{eq39}, $A_{k}$ is given as: 
\begin{align}     \label{eq-A-12}
\vert A_{k}^{\nu_{0}}\vert^{4}&\!\!=\!k^{2}\left[\!\left(\eta_{\nu_{0}}+\!i\zeta_{\nu_{0}}\!-\!\frac{1}{6}(\omega^{3}+(\omega^{*})^{3})\eta^{\prime\prime\prime}_{\nu_{0}}\right)\left(\mu\!-\!\frac{1}{2}k^{2}\eta^{\prime\prime}_{\nu_{0}}\right) \right. \nonumber \\
&\left. +\!\frac{1}{4}k^{2}\left(\eta^{\prime}_{\nu_{0}}\right)^{2}\!+\!\frac{1}{6}(\omega^{2}+(\omega^{*})^{2})k^{2}\eta^{\prime}_{\nu_{0}}\eta^{\prime\prime\prime}_{\nu_{0}}\!+\!\frac{1}{9}\vert\omega\vert^{4}k^{2}\left(\eta^{\prime\prime\prime}_{\nu_{0}}\right)^{2} \right].
\end{align}

\section*{Acknowledgements}

The authors do like to acknowledge the support of Graduate University of Advanced Technology for their support through their grant program. The authors aknowledge the support of Prof. Angelo Bassi which enabled AS to work with for nine months at the University of Trieste. One of us, AS, would like to thank the Ministry of Science, Research and Technology of Iran and the University of Trieste for their support that enabled him to spend nine months at the University of Trieste on a study leave.

\end{document}